\documentclass[conference,a4paper]{IEEEtran}

\usepackage[utf8]{inputenc} 
\usepackage[T1]{fontenc}
\usepackage{cancel}
\usepackage{url}
\usepackage{ifthen}
\usepackage{cite}
\usepackage{verbatim}
\usepackage{amssymb}
\usepackage[cmex10]{amsmath} 
\usepackage[cmex10]{mathtools} 
\usepackage{graphicx}
\usepackage{epstopdf}
\usepackage{subcaption}
\usepackage{units}
\usepackage{microtype}
\usepackage[hidelinks,
 			pdfauthor={Tommaso Catuogno, Menelaos Ralli Camara, Marco Secondini},
            pdftitle={Non-parametric Estimation of Mutual Information with Application to Nonlinear Optical Fibers},
            pdfsubject={Mutual Information Estimators},
            pdfkeywords={Mutual Information,Non-parametric estimation,Nonlinear Optical Fibers},
            pdfproducer={Menelaos Ralli Camara, Tommaso Catuogno},
            pdfcreator={none}]{hyperref}

 \graphicspath{{Figures/}}
 \DeclareGraphicsExtensions{.jpg,.eps}

\begin{document}
\title{Non-parametric Estimation of Mutual Information with Application to Nonlinear Optical Fibers} 


\author{%
  \IEEEauthorblockN{Tommaso Catuogno}
  \IEEEauthorblockA{Ericsson Research\\
                    Pisa, Italy\\
  with\\
  Scuola Superiore Sant'Anna,\\ 
  TeCIP institute\\
  at the time of this research\\\\
  tommaso.catuogno@ericsson.com\\
                    }
  \and
  \IEEEauthorblockN{Menelaos Ralli Camara}
  \IEEEauthorblockA{Scuola Superiore Sant'Anna,\\ 
  TeCIP institute\\
  Pisa, Italy
  \\\\
  menelaos.rallicamara@santannapisa.it}
  \and
  \IEEEauthorblockN{Marco Secondini}
  \IEEEauthorblockA{Scuola Superiore Sant'Anna,\\ 
  TeCIP institute\\
  Pisa, Italy
  \\\\
  marco.secondini@santannapisa.it}
}


\maketitle

\begin{abstract}
This paper compares and evaluates a set of non-parametric mutual information estimators with the goal of providing a novel toolset to progress in the analysis of the capacity of the nonlinear optical channel, which is currently an open problem. In the first part of the paper, the methods of the study are presented. The second part details their application to several optically-related channels to highlight their features.
\end{abstract}


\section{Introduction}
The evaluation of the capacity of the optical fiber channel is still an open problem, mainly due to the unavailability of an exact and mathematically tractable channel model \cite{agrell2016roadmap}.
There have been several studies on the fiber capacity limits (see, for instance, \cite{mitra:nature,essiambre2010capacity,Kramer2015, secondini2017scope} and references therein). In particular, many works have derived numerically-computed lower bounds (or their analytical approximations) implicitly or explicitly based on the use of an auxiliary-channel lower bound \cite{arnold2006simulation}. The main limit of this approach is that the tightness of the bound is determined by the accuracy with which the auxiliary channel approximates the true one, that is, by our knowledge of the true channel. Often, a simple Gaussian auxiliary channel is used (for numerical convenience and/or lack of a better knowledge) which, however, has been proved to give very loose bounds in some cases \cite{turitsyn2003information},\cite{agrell2014capacity},\cite{secondini2017scope}. On the other hand, the capacity of the optical fiber channel is upper bounded by the capacity of an equivalent additive white Gaussian noise (AWGN) channel with same total accumulated noise \cite{Kramer2015}. The gap between the tightest available lower and upper bounds is still unsatisfactorily large at high powers. 

In this work, we introduce a family of non-parametric mutual information (MI) estimators, based on k-nearest neighbour (kNN) statistics, which do not require any a priori information about the underlying channel model and which can be used to progress in the information theoretical analysis of the optical fiber channel. Other approaches, based for instance on the use of histograms, have been already explored \cite{fehenberger2014mutual}. The methods considered in this work are of particular interest since they combine the simplicity and adaptability of the histogram method with the peculiar  capability to work over multidimensional spaces, which is a key feature in the presence of memory. 

The paper is organized as follows. Section~II introduces three different non-parametric estimators based on kNN statistics, as well as the auxiliary-channel lower bound. Section~III and IV present the application of the methods to some classical simple channels and to some optical fiber channels, respectively. Conclusions are finally drawn in Section~V.

\section{Estimation of mutual information}
\subsection{Kozachenko estimator}

A possible approach to directly estimate the entropy of a random variable from its realizations is through kNN statistics. Given a continuous random vector $X\in\mathbb{R}^d$,  
its entropy is defined as 

\begin{equation}
H(X)= -E\lbrack\ \log(X)] 
\label{eq:Entropy_def}
\end{equation}
where the logarithm is in base 2, as all the others in the paper.
Let also $p(x)$ be the probability density function (pdf) of $X$ evaluated in $x$. Our goal is to estimate $H(X)$ from $N$ realizations $x_1,x_2,\ldots,x_N$ of the random vector $X$, without knowing $p(x)$.
The kNN approach, rather than estimating $p(x)$ from the available realizations (as typical histogram-based methods do), directly estimates  $\log p(x)$ from the statistics of the distance between $x$ and its $k$-th nearest neighbour in the available data set. The obtained estimator is described in \cite{kozachenko1987sample} and is named \textit{Kozachenko-Leonenko Entropy Estimator}, that is:

\begin{equation}
\hat{H}(X) = -\psi(k)+\psi(N)+\log c_d + \frac{d}{N}\sum_{i=1}^{N}\log\epsilon(i)
\label{eq:kozaH}
\end{equation}
where $c_d$ is the volume of the $d$-dimensional unit sphere and $\epsilon(i)$ is defined as the distance, over a chosen metric, between sample $x_i$ and its k-neighbour.
By using the same approach, it is also possible to estimate the mutual information $I(X;Y)$ between two random vectors, $X$ and $Y$. This is done by using the relation 
\begin{equation}
I(X,Y) = H(X) + H (Y) - H(X,Y)
\label{eq:mutual_info}
\end{equation}
and using (\ref{eq:kozaH}) to estimate the input, output, and joint entropies appearing in (\ref{eq:mutual_info}).

\subsection{Kraskov Estimator}
\label{Kraskov}

The Kozachenko estimator has also been used to develop another class of MI estimators that are more stable and works particularly well for low-dependent random variables \cite{kraskov2004estimating}. The obtained estimator, referred to as \textit{Kraskov Estimator},
differs from Kozachenko's one for the choice of the parameter $k$. In the latter, $k$ is the same for the estimation of the joint and marginal entropies, leading to a comparison of entropies evaluated on different scales. 

In contrast, in \cite{kraskov2004estimating}, a simple solution to bypass the problem is illustrated. Given a multidimensional plane, let $n_x(i)$ be the number of points in the interval $[x_i-\epsilon(i)/2,x_i+\epsilon(i)/2]$. The distance from $x_i$ to its $n_x(i)+1$ neighbor is equal to $\epsilon / 2$. Therefore, we can substitute $k$ with $n_x(i)+1$ in (\ref{eq:kozaH}), obtaining
\begin{equation}
\begin{split}
\hat{H}(X)\approx -\frac{1}{N}\sum_{i=i}^N  & \psi(n_x(i)+1) \\ 
& +\psi(N) +\log c_{dx} + \frac{d_x}{N}\sum_{i=1}^{N}\log\epsilon(i).
\end{split}
\label{eq:Kraskov_entropy}
\end{equation} 
By substituting (\ref{eq:Kraskov_entropy}) in (\ref{eq:mutual_info}) we eventually obtain the Kraskov estimator.

\subsection{Local Gaussian Estimator}

Although the previous estimators work well in low dimensions, they loose accuracy when the dimensionality of the random variable is high or the distribution highly non-uniform in the defined volumes. In fact, the primary source of errors in these methods is due to the assumption of a constant density in each volume, yielding---in cases of a highly concentrated probability mass function---an overestimate of the entropy. To overcome this problem, a different approach is to approximate the probability at sample $x_i$ by 
\begin{equation}
p(x)\approx \rho \exp(-\frac{1}{2}(x-\mu)^T \ S^{-1}  (x-\mu)  )
\label{eq:LocalGaussian1}
\end{equation}
where $\mu$ and $S^{-1}$ represent the empirical mean and covariance matrix of the $p$ neighbors of the point $x_i$. With this approach, the volume surrounding each sample is maintained constant, while the distribution within it is considered non-uniform, and in particular, it is assumed locally Gaussian with mean and covariance estimated on a set of points around the one under study. To obtain an equation of the same form as the previous methods, the volume related to sample $x_i$ is approximated by 
\begin{equation}
P_i= p(x_i) \frac{1}{g(x_i)} G_i
\label{eq:LocalGuassian2}
\end{equation}
where $G_i =  \int_{||x-x_i||<\epsilon/2} g(x) dx$   and  $g(x_i)=\exp(-\frac{1}{2}(x_i-\mu)^T \ S^{-1}  (x_i-\mu)  )$ .
Using (\ref{eq:LocalGaussian1}) and (\ref{eq:LocalGuassian2}) in the derivation of the Kozachenko Estimator, it is possible to define the \textit{Local Gaussian entropy estimator} \cite{lombardi2016nonparametric}
\begin{equation}
\hat{H}(x)=\psi(N)-\psi(k)-\frac{1}{N}\sum_{i=1}^N \log(g(x_i)) + \frac{1}{N}\sum_{i=1}^N \log G_i
\label{eq:LocalGuassian3}
\end{equation}
By comparing this estimator with the previous ones, it can be shown that, while close to the mode of the distribution the approximations to the integral of the probability density are similar among all the estimators, in the tails the integral is better captured by the Local Gaussian Estimator. This feature is particularly relevant in high-dimensional space distributions, as demonstrated in \cite{lombardi2016nonparametric}, where this method outperforms all the others.

\subsection{Auxiliary-channel lower bound}
A completely different approach to estimate the MI of a channel from a set of input and output realizations, without having an explicit knowledge of the underlying  model $p(y|x)$, is the use of the auxiliary-channel lower bound \cite{arnold2006simulation}
\begin{equation}
I(X;Y)\ge E\Big\{\log\frac{q(y|x)}{q(y)}\Big\}\simeq\frac{1}{N}\sum_{i=1}^{N}\log\frac{q(y_{i}|x_{i})}{q(y_{i})}
\label{eq:mutual_info_estimation}
\end{equation}
In (\ref{eq:mutual_info_estimation}), $q(y|x)$ and $q(y)=\int q(y|x)p(x)dx$ are, respectively, the conditional and output pdf of an arbitrarily selected auxiliary channel, while the expectation is taken with respect to the actual input--output joint distribution $p(x,y)$. The choice of the auxiliary channel affects the tightness and the computability of the bound (\ref{eq:mutual_info_estimation}), not its validity. Moreover, the bound is achievable by a mismatched detector that is optimized for the auxiliary channel.
This approach has been implicitly or explicitly adopted in optical fiber communications to obtain some capacity lower bounds (e.g., \cite{mitra:nature},\cite{essiambre2010capacity}). In many cases, a simple Gaussian-auxiliary-channel lower bound (GLB)---which is exact in the linear regime and, hence, accurate at low optical powers---has been considered, obtaining an easily computable bound that is achievable by conventional detectors \cite{secondini2013achievable}.
\section{Application to classical channels}

In this section, we consider two simple scenarios, for which the MI is exactly known, to investigate the accuracy of the considered MI estimators and their behaviour with respect to specific channel characteristics---namely, different signal-to-noise ratios and the presence of interference. 

\subsection{AWGN channel}

\begin{figure}[ht]
\centering
  \includegraphics[width=0.4\textwidth]{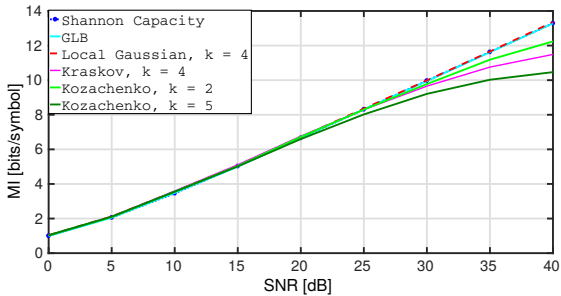}
 \caption{AWGN Channel - $N=15000$}
\label{fig:awgn_channel}
 \end{figure}

We start by considering an AWGN channel, $y = x + n$, and a circularly symmetric complex Gaussian input distribution. In this case, the input distribution is capacity achieving and the MI equals channel capacity $I(X;Y)=C=\log(1+\mathrm{SNR})$.
Fig.~(\ref{fig:awgn_channel}) compares the exact MI (capacity) with the estimates obtained through the approaches described in the previous section, considering a total number of samples $N= 15000$. In particular, the kozachenko estimator is computed in two flavors, with $k=2$ and $k=5$, to highlight the performance differences; Kraskov method is executed instead with $k=4$, the optimal parameter for this case.Finally the Local Gaussian estimator is ran with $k=4$ and $p= 0.04N$, a value suggested in the paper of reference for the method.
Regarding the auxiliary channel, in this simple case, is exactly matched to the true channel, such that the auxiliary-channel lower bound provides a very accurate estimate for any SNR. On the other hand, all the non-parametric methods based on kNN statistics become less accurate for high SNRs. The maximum SNR for which the MI can be accurately estimated in this example is between 25 and 30~dB, depending on the selected method and parameters. This is due to the increased correlation between the input and output samples, which causes an overestimation of the entropy while computing the volumes around the samples in the joint space. In principle, this limit can be increased at will by increasing the number of samples $N$, at the expense, however, of significantly increasing the computational effort.

\subsection{$2\times 2$ channel}
After testing the behaviour of the methods at different SNRs, we investigate their ability to estimate the MI in the presence of interference. To this end, we consider a $2\times 2$ multiple-input multiple-output (MIMO) channel $y=Hx+n$ in which the channel matrix
\begin{equation}
H=\left(\begin{array}{cc}
\cos\alpha & \sin\alpha\\
-\sin\alpha & \cos\alpha
\end{array}\right)
\label{eq:cross_channel}
\end{equation}
corresponds to a fixed rotation by an angle $\alpha$ and $n$ is a noise vector of two i.i.d. circularly symmetric complex Gaussian variables. For $\alpha=0$, we obtain two independent AWGN channels. Then, increasing $\alpha$, the information from the first (second) input is partly transferred also to the second (first) output, causing interference. Eventually, for $\alpha=\pi/2$, we have again two independent AWGN channels, with a crosswise input-output interconnection.

\begin{figure}[ht]
\centering
  \includegraphics[width=0.4\textwidth]{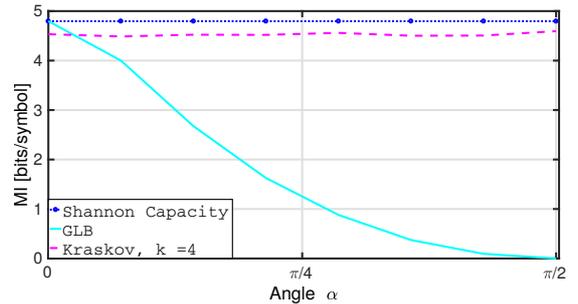}
 \caption{AWGN $2\times 2$ channel - $N = 11000$ SNR = 6dB}
\label{fig:doubleindependentchannel}
 \end{figure}

It is easy to verify that, regardless of the value of $\alpha$, the channel matrix is unitary and, hence, has no impact on the MI. In particular, for i.i.d. circularly symmetric complex Gaussian inputs, the MI equals the capacity of the channel, which is twice the capacity of the AWGN channel. However, if the channel model is unknown, it might be not straightforward to verify this result by observing only the channel realizations, unless a good non-parametric MI estimator is available. This is illustrated in Fig.~(\ref{fig:doubleindependentchannel}), in which the exact MI (capacity) is compared with the GLB and with the kNN Kraskov estimate for $\alpha$ in the range $[0,\pi/2]$ (a symmetric behaviour is obtained in $[\pi/2,\pi]$, and the curves have a period of $\pi$). The GLB is bound to measure the MI according to the underlying model specified by the selected auxiliary channel and has no ability to adapt to the specific characteristics of the true channel. Therefore, it is accurate only in the absence of any rotation and vanishes as the rotation approaches $\pi/2$. On the other hand, the Kraskov estimate is always close to the exact MI regardless of the value of $\alpha$ (a similar result is obtained with the other kNN-based estimators), showing a clear capability to "see" the MI between the input and output variables even when it is "mixed up" by an arbitrary rotation.

\section{Application to optical fiber channels}
In this section, we compare the estimators by considering three different scenarios related to the optical fiber channel---namely, a nonlinear channel with no dispersion, a linear dispersive channel, and a more realistic channel with both dispersion and nonlinearity.

\begin{figure*}[t!]
\centering
    \begin{subfigure}[t]{0.31\textwidth}
        \includegraphics[width=\textwidth]{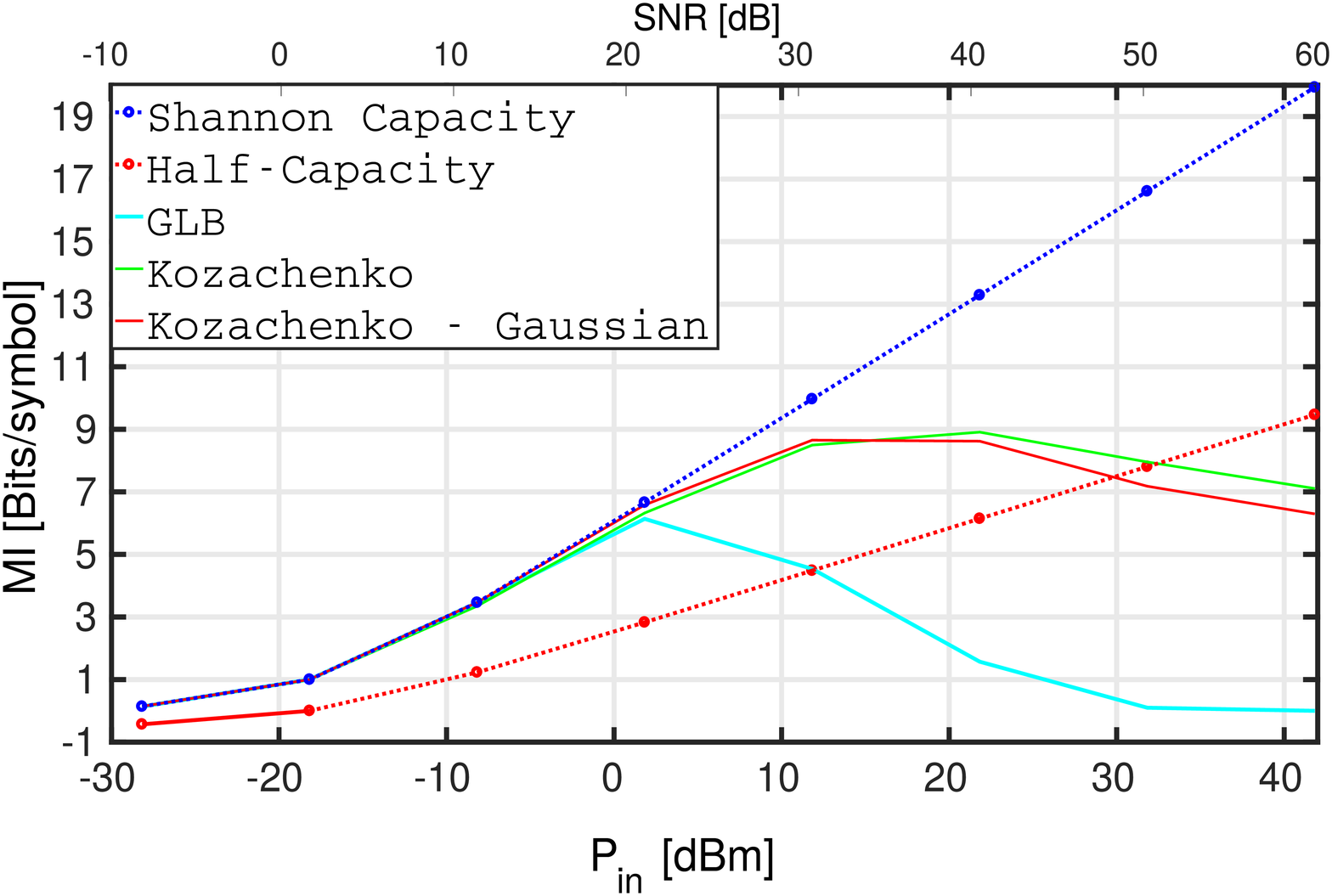}
        \caption{100 km $\times$ 12 span}
        \label{fig:100x12}
    \end{subfigure}
    ~ 
    \begin{subfigure}[t]{0.31\textwidth}
        \includegraphics[width=\textwidth]{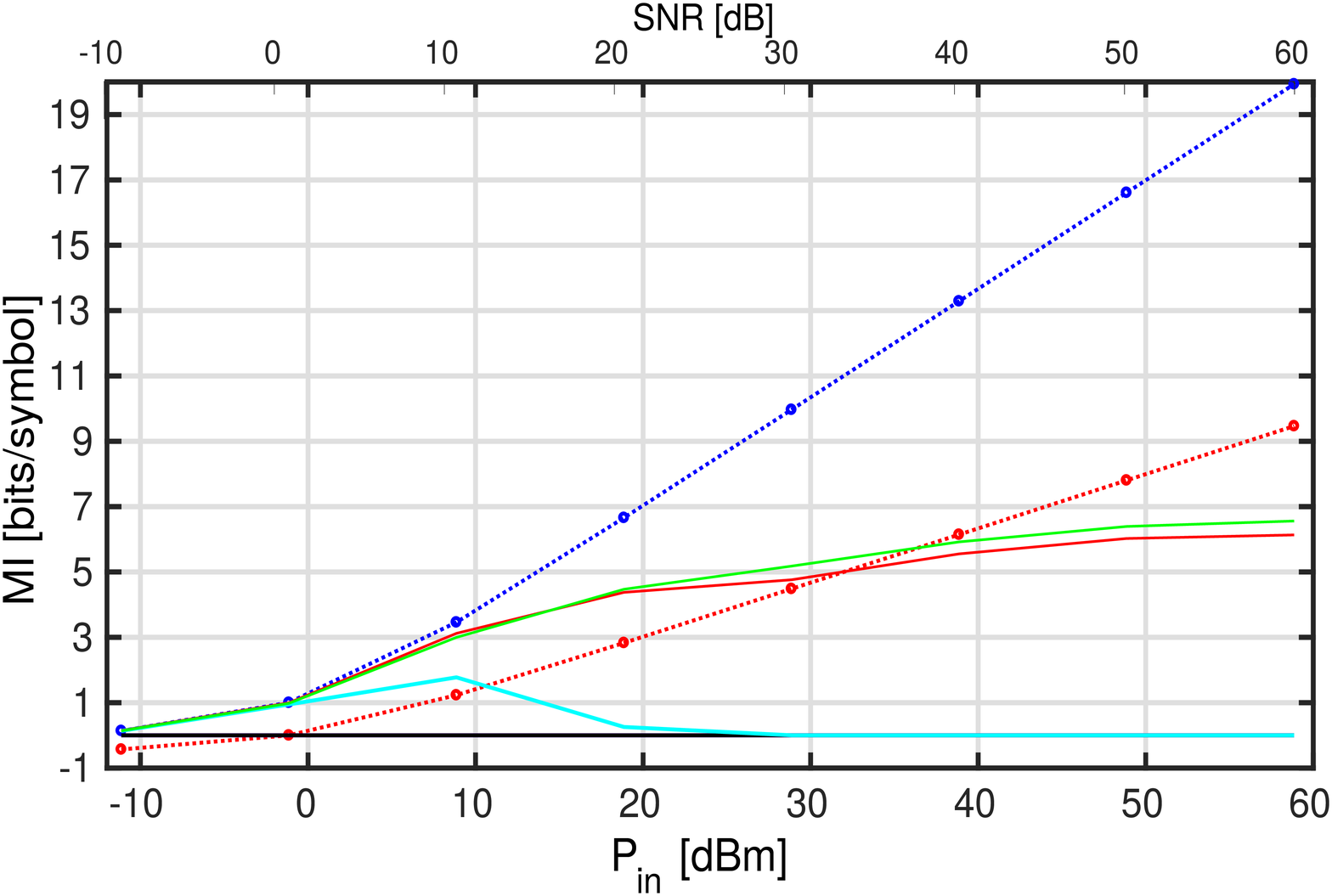}
        \caption{200 km $\times$ 6 span}
        \label{fig:200x6}
    \end{subfigure} 
    ~ 
    \begin{subfigure}[t]{0.31\textwidth}
        \includegraphics[width=\textwidth]{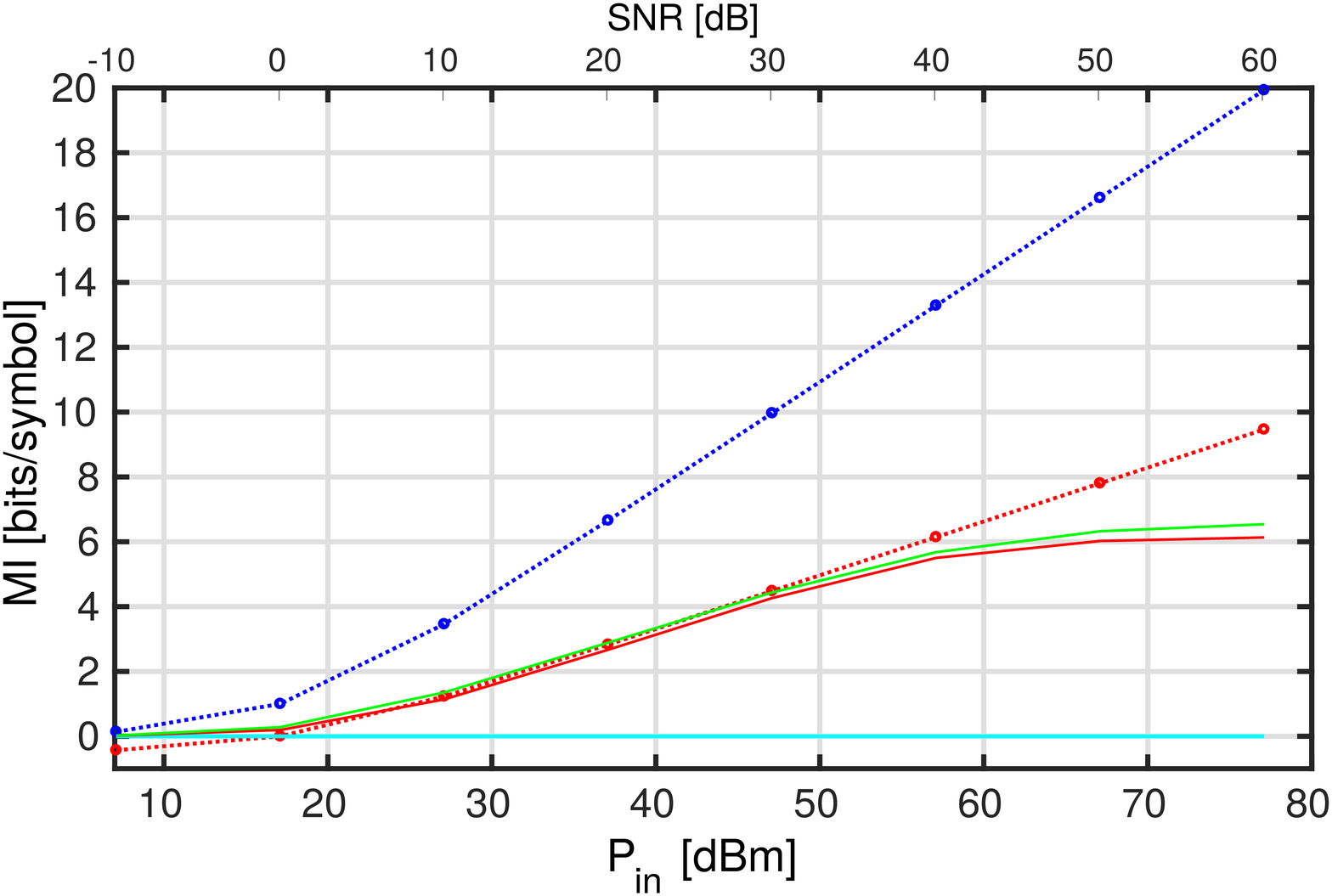}
        \caption{300 km $\times$ 4 span}
        \label{fig:300x4}
    \end{subfigure}
    \caption{Zero Dispersion Channel - $N = 2^{18}$}\label{fig:test1}
\end{figure*}

\subsection{Zero Dispersion}

The first channel under study is the zero-dispersion optical channel. In this case, the propagating signal is affected only by Kerr nonlinearity and amplified spontaneous emission (ASE) noise and the channel is memoryless. The capacity of this channel has been studied in  \cite{turitsyn2003information,5872133}.  When the input power $P_{\mathrm{in}}$ is low, the effect of nonlinearity is negligible, the channel is approximately AWGN, and its capacity approaches the AWGN channel capacity $C=\log(1+\mathrm{SNR})$. On the other hand, when $P_{\mathrm{in}}$ is large, the interaction between noise and Kerr nonlinearity generates a strong nonlinear phase noise that takes over the whole phase interval $[0,2\pi]$. In this case, the phase brings almost no information and the capacity is lower bounded as 
\begin{equation}
C \geq \frac{1}{2} \log(\mathrm{SNR})- \frac{1}{2} 
\label{eq:half-capacity}
\end{equation}
the bound being asymptotically exact for $P_{\mathrm{in}} \rightarrow \infty$ and achievable by an input distribution with half-Gaussian amplitude profile and uniform phase \cite{5872133}.

On the basis of such theoretical results, we tested the considered MI estimators to verify their accuracy in a high nonlinear regime, in which the conditional distribution of the channel may deviate significantly from a Gaussian distribution (see, for instance, \cite[Fig.~4]{5872133}). We considered three links with same total length of 1200~km but different span length and number (12x100, 6x200, and 4x300~km); a fiber attenuation $\alpha=\unit[0.2]{dB/km}$ and nonlinear parameter $\gamma=\unit[1.27]{W^{-1}km^{-1}} $; an amplifier noise figure of 6\,dB; and a symbol rate of 50\,GBd.
Fig.~\ref{fig:test1} shows the MI estimates as a function of the input power for the the three considered links, different estimators, and both a Gaussian and half-Gaussian input distribution. The linear capacity and the half-Gaussian capacity lower bound (\ref{eq:half-capacity}) are also plotted. In all the cases, the power range is selected to have a fixed SNR range, which is also reported (on the upper horizontal axis) as a reference. While for shorter spans the same SNR is obtained with a lower input power, therefore operating in a linear or weakly nonlinear regime, for longer spans it is achieved at a higher power, therefore operating in a strongly nonlinear regime.  
In the weakly nonlinear regime of Fig.~\ref{fig:100x12}, all the MI estimates are close to the linear capacity at low power, as expected. On the other hand, when the power is increased, the estimates reach a maximum and, then, decrease again. In the GLB case, this happens at significantly lower power (around 0~dBm) and is a typical behaviour of the estimator over the nonlinear optical channel \cite{secondini2017scope}. In fact, it is due to the mismatch between the true and auxiliary channel and cannot be modified by increasing the number of samples. On the other hand, the Kozachenko estimators are able to estimate a significantly higher MI at higher powers, reaching a peak at about 20\,dBm of input power. At very high power, the MIs fall below the lower bound, clearly indicating that also this (and the other kNN) methods are no longer accurate. This is not, however, an intrinsic limit of the method, but rather a limitation induced by the number of samples considered for the estimation, as already discussed in the AWGN case of Fig.~\ref{fig:awgn_channel}. In fact, the maximum estimated MI and the crossing point with the capacity lower bound depend on  $N$ and can be increased by increasing it.

The difference between the GLB and the kNN estimator is even more evident in cases (b) and (c), in which a strongly nonlinear regime is already achieved at lower SNR and MI. While the GLB is able to measure very little or almost  zero MI, the Kozachenko estimator gradually decreases from the linear capacity to the half-capacity lower bound as the power increases.  

\subsection{Dispersive Channel}

As a second test, we consider the case of a linear dispersive optical channel, with the goal of analyzing the impact of channel memory on the estimators.
A linearly-modulated 10\,GBd signal is propagated through a link consisting of a few (from 1 to 5) spans of 80\,km \textit{single mode fiber} (SMF)---with dispersion coefficient $D=\unit[16]{ps/nm/km}$ and attenuation coefficient $\alpha=\unit[0.2]{dB/km}$---each followed by an optical amplifier with a noise figure of 5.3\,dB. The pulse shape has a root-raised-cosine Fourier transform with rolloff factor 0.2, and i.i.d. circularly symmetric complex Gaussian input symbols are considered. At the receiver, after matched filtering and symbol time sampling, no other processing is performed, such that a significant inter-symbol interference (ISI) due to chromatic dispersion is present.

\begin{figure}[ht]
\centering
  \includegraphics[width=0.4\textwidth]{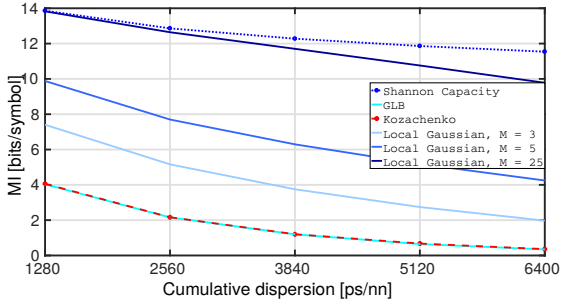}
 \caption{Dispersive Channel - $N=2^{18}$}
\label{fig:dispersion}
 \end{figure}

Analogously to the case of the MIMO channel investigated in the previous section, dispersion corresponds to a unitary transformation. Therefore, it does not affect the MI, which in this case equals the capacity of the AWGN channel. However, without an explicit knowledge of the channel model, it might be hard to extract the information from the received samples, being it spread and mixed up over several received symbols. This is clearly apparent from Fig.~\ref{fig:dispersion}, if the GLB is considered, which simply neglects channel memory and corresponds to a symbol-by-symbol detection. In this case, the estimated MI is much lower than the actual one, and rapidly vanishes as the accumulated dispersion (number of spans) increases. A very similar behaviour is obtained when considering the Kozachenko estimator, still on a symbol-by-symbol basis. On the other hand, we should be able to fully recover the information from the output samples if they are jointly processed. To this end, we consider the local-Gaussian estimator (which is known to be more robust in a high-dimensional space) and estimate the MI between each input sample and a block of 3, 5, or 21 output samples. As expected, the MI estimates increase when increasing the block size, as a longer portion of the channel memory is considered. In particular, for 21 output samples, the estimate approaches the exact value after a single span, and remains reasonably high for longer distances, clearly showing a good behaviour in a high-dimensional space (two real input variables and 42 real output variables) and a significant ability to cope with severe ISI.

\subsection{Realistic optical channel}

\begin{figure}[ht]
\centering
  \includegraphics[width=0.4\textwidth]{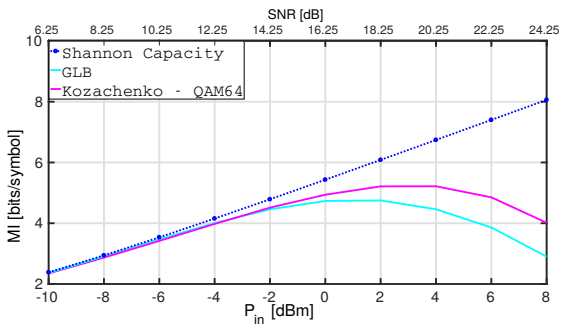}
 \caption{Realistic Channel - $N=2^{18}$}
\label{fig:realistic}
 \end{figure}

As a final test, we consider a realistic optical channel with both dispersion and nonlinearity. Moreover, to test the ability of the estimators to work also with discrete distributions, we consider a 64QAM modulation. The considered scenario is the same described in \cite{irukulapati2016tighter}. The symbol rate is 14\,GBd, and a link of 30x120\,km spans of SMF is considered, with $\alpha=\unit[0.2]{dB/km}$, $D=\unit[16]{ps/nm/km}$, $\gamma=[1.3]{W^{-1} km^{-1}}$. Each span is followed by an optical amplifier with a noise figure of $5.5dB$ and an ideal fiber Bragg grating that exactly compensates for dispersion. At the output, digital backpropagation (DBP) is used to mitigate linear and nonlinear ISI (being though ineffective against signal-noise interaction), after which matched filtering and symbol-time sampling are performed. Fig. (\ref{fig:realistic}) compares the GLB and the Kozachenko estimate with the capacity of the linear channel. With respect to the GLB, the Kozachenko estimator is able to cope with the non-Gaussian distribution induced by signal-noise interaction and give a higher MI estimate. Remarkably, the Kozachenko estimate equals the tightest MI lower bound obtained in \cite{irukulapati2016tighter} by employing a complex algorithm---stochastic DBP with Gaussian message passing---specifically designed to include the effect of signal-noise interaction in DBP. 

\section{Conclusion}
We have explored some non-parametric approaches for the estimation of the MI, based on kNN statistics. This approaches are especially useful when no knowledge of the channel model is available, as in the case of the optical fiber channel, about which very little is known in terms of capacity limits. The considered methods have been tested in several different scenarios---including simple classical channel models and more realistic optical fiber channels. They are accurate for a wide range of SNRs, and are able to cope with non-Gaussian distribution, interference, and a long channel memory. In all the considered cases, the non-parametric methods have outperformed the Gaussian-auxiliary-channel lower bound and approached either the exact MI (when available) or the tightest available bounds, never exceeding them.

These encouraging results suggest that kNN-based non-parametric MI estimators might be a useful tool for the analysis of the optical fiber channel and deserve more attention. Further investigations are required to verify the effectiveness of the methods in different scenarios, such as wavelength-division multiplexing systems, and possibly to build provable lower bounds out of these methods.





\bibliographystyle{IEEEtran}
\bibliography{bibliography.bib}

\end{document}